\documentclass[aps, preprint, showpacs, superscriptaddress, nofootinbib,
amsmath, amssymb]{revtex4}

\usepackage{graphicx}

\DeclareMathOperator{\Tr}{Tr}

\newcommand{\order}[1]{\mathcal{O}\left(\epsilon^{#1}\right)}

\def\bwt{\begin{widetext}}
\def\ewt{\end{widetext}}
\def\be{\begin{equation}}
\def\ee{\end{equation}}
\def\bea{\begin{eqnarray}}
\def\eea{\end{eqnarray}}
\def\bean{\begin{eqnarray*}}
\def\eean{\end{eqnarray*}}
\def\bary{\begin{array}}
\def\eary{\end{array}}
\def\bit{\begin{itemize}}
\def\eit{\end{itemize}}

\def\su5u1{SU(5) \times U(1)}
\def\fsu5u1{SU(5) \times U(1)'}
\def\so10{SO(10)}
\def\sq20{SO(10) \times SO(10)}

\begin{document}

\title{Fermion Masses and Mixings in GUTs with Non-Canonical $U(1)_Y$}

\author{Ilia Gogoladze}
\email{ilia@physics.udel.edu}
\affiliation{Bartol Research Institute, Department of Physics and Astronomy,
University of Delaware, Newark, DE 19716, USA}

\author{Chin-Aik Lee}
\email{jlca@udel.edu}
\affiliation{Bartol Research Institute, Department of Physics and Astronomy,
University of Delaware, Newark, DE 19716, USA}

\author{Tianjun Li}
\email{tli@itp.ac.cn}

\affiliation{George P. and Cynthia W. Mitchell Institute for
Fundamental Physics, Texas A$\&$M University, College Station, TX
77843, USA }

\affiliation{Institute of Theoretical Physics, Chinese Academy of Sciences,
 Beijing 100080, P. R. China}

\author{Qaisar Shafi}
\email{shafi@bartol.udel.edu}
\affiliation{Bartol Research Institute, Department of Physics and Astronomy,
University of Delaware, Newark, DE 19716, USA}

\date{\today}

\begin{abstract}


We discuss fermion masses and mixings in models derived from orbifold GUTs such
that gauge coupling unification is achieved without low energy supersymmetry by
utilizing a non-canonical $U(1)_Y$. A gauged $U(1)_X$ flavor symmetry plays an
essential role, and the Green-Schwarz mechanism is invoked in anomaly
cancellations. Models containing vector-like particles with masses close to
$M_{\text{GUT}}$ are also discussed.

\end{abstract}


\preprint{BA-07-20, MIFP-07-12}

\maketitle

\section{Introduction}


In some recent papers \cite{Barger:2005gn,Gogoladze:2006ps}, it has
been shown that unification of the Standard Model (SM) gauge
couplings can be realized without invoking low energy supersymmetry
(SUSY). The unification scale turns out to be close to $M_\text{GUT}
\sim 4 \times 10^{16}$ GeV. Such models are realized from
supersymmetric $SU(N)$ gauge theories in higher dimensions
compactified on suitable orbifolds, and showing that the
normalization of $U(1)_Y$ can be different from the standard value
of $5/3$. One well known example utilizes a value of $4/3$ for the
hypercharge normalization, with SUSY broken at $M_\text{GUT}$.  An
important extension of these ideas implements gauge and Yukawa
coupling unification at $M_\text{GUT}$ \cite{Gogoladze:2006ps}. For
instance, with gauge-top quark Yukawa coupling unification and with
SUSY broken at $M_{\text{GUT}}$, the SM Higgs boson mass turns out
to be $135\pm 6$ GeV \cite{Gogoladze:2006ps}. Somewhat larger
values for the Higgs mass, $144\pm 4$ GeV, are found with gauge-bottom
quark Yukawa coupling unification \cite{Gogoladze:2006ps}. Examples
based on split SUSY have also been discussed
\cite{Gogoladze:2006qp}. For a discussion of models with low energy SUSY
and non-canonical normalization of $U(1)_Y$, see
\cite{Gogoladze:2006si}.

Our main goal here is to understand the SM fermion masses and mixings in this
framework by employing a flavor $U(1)_X$ symmetry via the Froggatt-Nielsen
(FN) mechanism~\cite{Froggatt:1978nt}.
The $U(1)_X$ symmetry is gauged and has anomalies which are cancelled by the
Green-Schwarz (GS) mechanism~\cite{MGJS}.
This avoids potential problems which may arise if
a global $U(1)_X$ symmetry is employed to implement the FN mechanism. Because
the $U(1)_X$ anomaly cancellations depend on the normalization of $U(1)_Y$, and
our models exhibits SUSY at $M_{\text{GUT}}$, some care is required to realize
a consistent framework for the SM fermion masses and mixings in our setup. For
viable
fermion textures, we follow closely the discussions presented in Refs.
\cite{Dreiner:2003hw,Dreiner:2003yr,Dreiner:2006xw}. One of the two
scenarios includes vector-like particles with
masses close to $M_{\text{GUT}}$.

The paper is organized as follows. In Section \ref{sec:review}, we briefly
review the FN mechanism realized with an anomalous $U(1)_X$. The simplest models
are discussed in Section \ref{sec:no_extra}, while Section \ref{sec:additional}
contains models with additional vector-like particles. Our conclusions are
summarized in Section \ref{sec:conclusion}.

\section{FN mechanism and Yukawa textures}
\label{sec:review}

The SM fermion masses and mixings can be explained elegantly via
the FN mechanism~\cite{Froggatt:1978nt} where an additional
 flavor dependent global $U(1)_X$ symmetry is introduced.
To stabilize this mechanism against
quantum gravity corrections,  we consider an
anomalous gauged $U(1)_X$ symmetry.
In a weakly coupled heterotic string theory, there exists an
anomalous $U(1)_X$ gauge symmetry where the corresponding
anomalies are cancelled by the GS mechanism~\cite{MGJS}.
For completeness, let us briefly review gauge coupling unification and
anomaly cancellations in weakly coupled heterotic string
model building.

The generic four-dimensional
Lagrangian in the Einstein metric from four-dimensional
string theories can be written as
\begin{align}
\mathcal{L} &\supset \int d^4x \int d^2\theta
S \left\{ \frac{k_C}{2} \Tr_C\left[W_C^\alpha W_{C\alpha}\right] +
\frac{k_W}{2}
\Tr_W\left[W_W^\alpha W_{W\alpha}\right] + \right.\nonumber\\
&\quad\left. \frac{k_{Y}}{4}  W_Y^\alpha W_{Y\alpha} + \frac{k_{X}}{4}
W_X^\alpha W_{X\alpha} + \frac{k_{XY}}{2} W_Y^\alpha W_{X\alpha} \right\} +
{\rm H. C.} ~,~\,
\label{SFF-gauginomass}
\end{align}
where we neglect the overall constant here.
The last term is the gauge kinetic mixing term between $U(1)_Y$
and $U(1)_X$. $k_C$, $k_W$, $k_Y$, $k_X$ and $k_{XY}$ are the
Kac-Moody levels for the gauge symmetries $SU(3)_C$, $SU(2)_L$, $U(1)_Y$,
$U(1)_X$ and the $U(1)_Y\times U(1)_X$ gauge kinetic
mixing term, respectively. $k_C$ and $k_W$ are positive integers,
$k_Y$ and $k_X$ are positive rational numbers,
and $k_{XY}$ is a rational number.
 Also, the generators for the non-Abelian gauge groups are
normalized to $1/2$ for the fundamental representation.
With $g^{-2}_S$ denoting the vacuum expectation value (VEV)
 of the real part of $S$, the
gauge coupling unification~\cite{Ginsparg:1987ee, Dienes:1996du} is given by
\begin{equation}
k_C g_C^2=k_W g_W^2=k_{Y}g_{Y}^2=k_{X}g_{X}^2=k_{XY}g_{XY}^2=g_S^2~,~\,
\end{equation}
where $g_C$, $g_W$, $g_Y$, $g_X$, and $g_{XY}$ are the gauge couplings
for $SU(3)_C$, $SU(2)_L$, $U(1)_Y$, $U(1)_X$ gauge symmetries and
the $U(1)_Y\times U(1)_X$ gauge kinetic
mixing term, respectively.
In addition, there exist the following terms for the
imaginary part of $S$ (${\rm Im}(S)$) and the gauge field strengths
\begin{align}
\mathcal{L} &\supset \int d^4x
{\rm Im}(S) \left\{ \frac{k_C}{2} \Tr_C\left[F_C \wedge F_C\right] +
\frac{k_W}{2} \Tr_W\left[F_W \wedge F_W\right]
+ \right.\nonumber\\
&\quad\left. \frac{k_{Y}}{4} F_Y \wedge F_Y + \frac{k_{X}}{4} F_X \wedge F_X
+   \frac{k_{XY}}{2} F_X \wedge F_Y \right\} +
{\rm H. C.} ~.~\,
\label{Eq-AA1}
\end{align}
The NS-NS
 two-form field $B_{\mu \nu}$ in  four dimensions can couple
to the $F_X$ as follows:
\begin{equation}
\mathcal{L} ~\supset~ \int d^4x \:c_0\: B \wedge F_X~,~\,
\label{Eq-AA2}
\end{equation}
where $c_0$ is a constant.
Because $B_{\mu \nu}$ and ${\rm Im}(S)$ are dual to each
other in four dimensions, we obtain that
Eqs. (\ref{Eq-AA1}) and (\ref{Eq-AA2}) can contribute
to the gauge anomalies after we contract $B_{\mu \nu}$ and ${\rm Im}(S)$.
Therefore, the total anomaly cancellations give us the following
constraints~\cite{Ginsparg:1987ee, Dienes:1996du}
\begin{equation}
\label{eq:Green_Schwarz}
\frac{\mathcal{A}_{CCX}}{k_C}=\frac{\mathcal{A}_{WWX}}{k_W}=
\frac{\mathcal{A}_{YYX}}{k_{Y}}
=\frac{\mathcal{A}_{XXX}}{3k_{X}}
=\frac{\mathcal{A}_{YXX}}{k_{YX}}
= \frac{\mathcal{A}_{GGX}}{24}~,~
\end{equation}
where $\mathcal{A}_{CCX}$, $\mathcal{A}_{WWX}$, $\mathcal{A}_{YYX}$,
$\mathcal{A}_{XXX}$, and $\mathcal{A}_{YXX}$ are the gauge anomalies
from the chiral fermions, and $\mathcal{A}_{GGX}$ is gauge-gravity
mixed anomaly from the chiral fermions. The last equality is required to cancel
the gauge-gravity mixed
anomaly.

We follow the standard notation for the SM left-handed quark doublets,
right-handed up-type quarks, right-handed down-type
quarks, left-handed lepton doublets, right-handed
neutrinos, and right-handed leptons by denoting them as
$q_i$, $u^c_i$, $d^c_i$, $l_i$, $\nu^c_i$, and $e^c_i$, respectively.
To reproduce the observed neutrino masses and mixings, we need
at least two, or even three right-handed neutrinos for the
seesaw mechanism~\cite{Seesaw}.
In the former case, one of the left-handed neutrinos will
remain massless, which is consistent
with the known data. To be concrete, we will consider
three right-handed neutrinos in this paper. Moreover,
since we consider GUT-scale SUSY breaking, there is one pair of
Higgs doublets $H_u$ and $H_d$ close to
the string scale.

To break the $U(1)_X$ gauge symmetry, we introduce a flavon field
$A$ with $U(1)_X$ charge $-1$. To preserve SUSY
close to the string scale, $A$ can acquire a VEV
so that the $U(1)_X$ D-flatness can
be realized.  It was shown~\cite{Dreiner:2003hw, Dreiner:2003yr} that
\begin{equation}
0.171 \le \epsilon\equiv{\frac{\langle A \rangle}{M_{\rm Pl}}} \le 0.221
~,~\,
\end{equation}
where $M_{\rm Pl}$ is the reduced Planck scale.
Interestingly, $\epsilon$ is about the size of the Cabibbo angle.
Note that the $X$ charges of the SM fermions and the
Higgs fields are denoted by appropriate subscripts.

With SUSY broken around $M_{\text{GUT}}$,
the SM fermion Yukawa couplings arising from
the holomorphic superpotential at the string scale are given by
\begin{eqnarray}
- \mathcal{L} &=& y^U_{ij} \left({\frac{A}{M_{\rm Pl}}}\right)^{XYU_{ij}}
q_i u^c_j H_u+
y^D_{ij} \left({\frac{A}{M_{\rm Pl}}}\right)^{XYD_{ij}} q_i d^c_j H_d
 \nonumber\\ &&
+y^E_{ij} \left({\frac{A}{M_{\rm Pl}}}\right)^{XYE_{ij}} l_i e^c_j H_d+
y^{N}_{ij} \left({\frac{A}{M_{\rm Pl}}}\right)^{XYN_{ij}} l_i \nu^c_j
H_u~,~\,
\end{eqnarray}
where $y^U_{ij}$, $y^D_{ij}$, $y^E_{ij}$, and $y^{N}_{ij}$ are
order one Yukawa couplings, and $XYU_{ij}$, $XYD_{ij}$, $XYE_{ij}$
and $XYN_{ij}$ are non-negative integers:
\begin{eqnarray}
&& XYU_{ij} = X_{q_i} + X_{u^c_j} + X_{H_u}~,~~
XYD_{ij} = X_{q_i} + X_{d^c_j} + X_{H_d}~,~\\
&& XYE_{ij} = X_{l_i} + X_{e^c_j} + X_{H_d}~,~
XYN_{ij} = X_{l_i} + X_{\nu^c_j} + X_{H_u}~.~\,
\end{eqnarray}

Our goal is to generate Yukawa textures that
can explain the SM fermion masses and mixings.
The quark textures $A$, $B$, and $C$ in Table~\ref{tab:textures},
with $\epsilon \approx 0.2$, as copied from
\cite{Dreiner:2003hw,Dreiner:2003yr,Dreiner:2006xw}, can reproduce the
SM quark Yukawa couplings and the CKM quark mixing matrix.
\begin{table}
\caption{The three quark textures.}
\label{tab:textures}
\[
\begin{array}{||c||c|c|c||}
\hline\hline
\text{Yukawa}   & A     & B     & C\\
\hline\hline
Y^U     &
\begin{pmatrix}\epsilon^{8}&\epsilon^{5}&\epsilon^{3}\\
\epsilon^{7}&\epsilon^{4}&\epsilon^{2}\\
\epsilon^{5}&\epsilon^{2}&\epsilon^{0}\end{pmatrix}     &
\begin{pmatrix}\epsilon^{8}&\epsilon^{6}&\epsilon^{4}\\
\epsilon^{6}&\epsilon^{4}&\epsilon^{2}\\
\epsilon^{4}&\epsilon^{2}&\epsilon^{0}\end{pmatrix}     &
\begin{pmatrix}\epsilon^{8}&\epsilon^{4}&\epsilon^{2}\\
\epsilon^{8}&\epsilon^{4}&\epsilon^{2}\\
\epsilon^{6}&\epsilon^{2}&\epsilon^{0}\end{pmatrix}\\
\hline
Y^D     &
\epsilon^c\begin{pmatrix}\epsilon^{4}&\epsilon^{3}&\epsilon^{3}\\
\epsilon^{3}&\epsilon^{2}&\epsilon^{2}\\
\epsilon^{1}&\epsilon^{0}&\epsilon^{0}\end{pmatrix}     &
\epsilon^c\begin{pmatrix}\epsilon^{4}&\epsilon^{4}&\epsilon^{4}\\
\epsilon^{2}&\epsilon^{2}&\epsilon^{2}\\
\epsilon^{0}&\epsilon^{0}&\epsilon^{0}\end{pmatrix}     &
\epsilon^c\begin{pmatrix}\epsilon^{4}&\epsilon^{2}&\epsilon^{2}\\
\epsilon^{4}&\epsilon^{2}&\epsilon^{2}\\
\epsilon^{2}&\epsilon^{0}&\epsilon^{0}\end{pmatrix}\\
\hline\hline
\end{array}
\]
\end{table}
And the following lepton textures can reproduce the neutrino masses and PMNS
neutrino mixing matrix:
\begin{align}
Y^E &\sim \epsilon^c\begin{pmatrix}\epsilon^{4}&\epsilon^{3}&\epsilon^{1}\\
\epsilon^{3}&\epsilon^{2}&\epsilon^{0}\\
\epsilon^{3}&\epsilon^{2}&\epsilon^{0}\end{pmatrix}~,~\,
\label{eq:YE}\\
M_{LL} &\sim \frac{\langle H_u \rangle^2}{M_s}
\epsilon^{-5}\begin{pmatrix}\epsilon^{2}&\epsilon^{1}&\epsilon^{1}\\
\epsilon^{1}&\epsilon^{0}&\epsilon^{0}\\
\epsilon^{1}&\epsilon^{0}&\epsilon^{0}\end{pmatrix}~,~\,
\label{eq:neutrino}
\end{align}
where $c$ is either 0, 1, 2 or 3, and
$\tan \beta \equiv \langle H_u \rangle / \langle
H_d \rangle$ satisfies $\epsilon^c \sim \epsilon^3 \tan \beta$. This neutrino
texture requires some amount of fine-tuning as it generically predicts
\begin{eqnarray}
\sin \theta_{12} \sim \epsilon  ~,~~
\Delta m^2_{12} \sim \Delta m^2_{23}~.~\,
\end{eqnarray}
Interestingly, with $\epsilon$ as large as $0.2$, the
amount of fine-tuning needed is
not that huge and this is shown in the computer simulations of
\cite{Dreiner:2003hw,Dreiner:2003yr,Dreiner:2006xw} with random values for the
coefficients.

To obtain the desired textures,
the $X$ charge assignments for the SM fermions and Higgs fields
 satisfy the following equations
\begin{subequations}
\begin{align}
X_{H_u} + X_{q_3} + X_{u^c_3} &= 0~,~\, \\
X_{q_2} - X_{q_3} &= 2~,~\, \\
X_{q_1} - X_{q_3} &= 3,4,2~,~\, \\
X_{u^c_2} - X_{u^c_3} &= 2~,~\, \\
X_{u^c_1} - X_{u^c_3} &= 5,4,6~,~\, \\
X_{H_d} + X_{q_3} + X_{d^c_3} &= c~,~\, \\
X_{d^c_2} - X_{d^c_3} &= 0~,~\, \\
X_{d^c_1} - X_{d^c_3} &= 1,0,2~,~\, \\
X_{H_d} + X_{l_3} + X_{e^c_3} &= c~,~\, \\
X_{l_2} - X_{l_3} &= 0~,~\, \\
X_{l_1} - X_{l_2} &= 1 ~.~\,
\end{align}
\end{subequations}
Finally, regardless of whether we use texture $A$, $B$
or $C$, the following equations always hold:
\begin{subequations}
\begin{align}
3X_{H_u} + \sum_i \left[ X_{q_i}+ X_{u^c_i} \right]&=12~,~\, \\
3X_{H_d} + \sum_i \left[ X_{q_i}+ X_{d^c_i} \right]&=6+3c~,~\, \\
3X_{H_d} + \sum_i \left[ X_{l_i}+ X_{e^c_i} \right]&=6+3c~.~\,
\end{align}
\end{subequations}

\section{Models without Extra Vector-Like Particles}
\label{sec:no_extra}

In this paper we assume that $k_C=k_W=1$ because the
semi-realistic string models have so far only been constructed
at the Kac-Moody level one for the non-Abelian gauge factors.
Due to the GUT-scale SUSY breaking,
we only have the SM as the effective theory
below $M_{\text{GUT}}$. It has been shown~\cite{Barger:2005gn} that gauge
coupling unification can be achieved if we choose
\begin{equation}
k_Y=\frac{4}{3}~,~~  k_C=k_W=1~.~\,
\label{eq:unification}
\end{equation}
Thus, we shall work with $k_C=k_W=1$.

Before we calculate the gauge anomalies, let us explain
our conventions. We define  the anomalous contributions
of the chiral fermions as follows:
\begin{align}
\mathcal{A}_{ABC} &= \frac{1}{2}\Tr_{\text{matter}}[T_A T_B T_C]~,~\,
\end{align}
for Abelian symmetries $A$, $B$ and $C$, and
\begin{align}
\mathcal{A}_{AAC} \frac{\Tr_A[T^a_A T^b_A]}{2} &=
\frac{1}{2}\Tr_{\text{matter}}[T^a_A
T^b_A T_C]~,~\,
\end{align}
for a non-Abelian symmetry $A$ and an Abelian symmetry $C$, where $\Tr_A$ is the
trace over a fundamental representation of $A$.
More specifically, the anomalies are given as follows
\begin{subequations}
\begin{align}
\mathcal{A}_{CCX}&= \sum_i \left[ 2X_{q_i} + X_{u^c_i} + X_{d^c_i} \right] ~,~\\
\mathcal{A}_{WWX}&= \sum_i \left[ 3X_{q_i} +
X_{l_i}\right]+ X_{H_u} + X_{H_d} ~,~\\
\mathcal{A}_{YYX}&= \sum_i \left[ \frac{1}{3}X_{q_i} +
\frac{8}{3}X_{u^c_i} + \frac{2}{3}X_{d^c_i} + X_{l_i} +
2X_{e^c_i}\right]+ X_{H_u} + X_{H_d} ~,~\\
\mathcal{A}_{YXX}&= \sum_i \left[ X_{q_i}^2 -
2X_{u^c_i}^2 + X_{d^c}^2 - X_{l_i}^2 + X_{e^c_i}^2  \right]
+ X_{H_u}^2 - X_{H_d}^2 ~,~\\
\mathcal{A}_{XXX}&= \sum_i \left[ 3X_{q_i}^3 + \frac{3}{2}X_{u^c_i}^3 +
\frac{3}{2}X_{d^c_i}^3 + X_{l_i}^3 + \frac{1}{2}X_{e^c_i}^3 +
\frac{1}{2}X_{\nu^c_i}^3 \right] \nonumber\\ & \quad
+X_{H_u}^3-X_{H_d}^3   - \frac{1}{2} + \dots ~,~ \\
\mathcal{A}_{GGX}&= \sum_{\phi} X_{\phi} ~,~\,
\end{align}
\label{eq:anomalies}
\end{subequations}
where $\mathcal{A}_{XXX}$ and $\mathcal{A}_{GGX}$ also
receive contributions from the
flavinos and possibly other superfields as well.

The gauge and gauge-gravity mixed anomalies can
be cancelled by the GS mechanism,
as discussed in the previous Section.
We will not consider $\mathcal{A}_{XXX}$
and $\mathcal{A}_{GGX}$, since  these anomalies can always be cancelled by
introducing additional SM singlet superfields which
 are charged under $U(1)_X$.
These extra superfields can be  decoupled
from the SM after $U(1)_X$ breaking,
which is why we may neglect them.
However, we would like to point out that
to explain the SM fermion masses and mixings via
the FN mechanism in $U(1)_X$ models,
the anomalies $\mathcal{A}_{XXX}$ and $\mathcal{A}_{GGX}$
are typically on the order of a thousand (${\cal O} (1000)$)!
To cancel these anomalies, we have to introduce literally
hundreds of SM singlets with suitable $U(1)_X$ charges.
It remains to be seen whether such $U(1)_X$ models
can arise from string theory.

We will now combine Eq. \eqref{eq:Green_Schwarz} with above
equations to solve for the $X$ charges. First, we can derive
the following equations:
\begin{align}
\mathcal{A}_{CCX}&=-3\left( X_{H_u} + X_{H_d}\right) + \left\{ 3X_{H_u} +
\sum_i \left[
X_{q_i}+ X_{u^c_i} \right] \right\} +\nonumber\\
& \quad \left\{
3X_{H_d} + \sum_i \left[ X_{q_i}+ X_{d^c_i} \right] \right\}\nonumber\\
&=-3\left( X_{H_u} + X_{H_d}\right) + 18+3c~,~\,
\end{align}

\begin{align}
\left(k_Y-\frac{5}{3}\right)\mathcal{A}_{CCX}&=
\mathcal{A}_{WWX}+\mathcal{A}_{YYX}-\frac{8}{3}\mathcal{A}_{CCX}
\nonumber\\
&=2\left( X_{H_u} + X_{H_d} \right) + \sum_i \left[
-2X_{q_i} -2X_{d^c_i} +2X_{l_i} + 2X_{e^c_i} \right]\nonumber\\
&=2\left( X_{H_u} + X_{H_d} \right) + \left\{ 3X_{H_d} + \sum_i \left[
X_{l_i}+ X_{e^c_i} \right] \right\} - \nonumber\\
& \quad \left\{ 3X_{H_d} +
\sum_i \left[ X_{q_i}+ X_{d^c_i} \right] \right\}\nonumber\\
&= 2\left( X_{H_u} + X_{H_d} \right)~,~\,
\end{align}

\begin{align}
\left(k_Y-\frac{5}{3}\right) \left\{ -3\left( X_{H_u} + X_{H_d}\right) + 18+3c
\right\} &= 2\left( X_{H_u} + X_{H_d} \right)~,~\, \nonumber\\
X_{H_u} + X_{H_d} &= \frac{\left(k_Y-\frac{5}{3}\right)(6+c)}{k_Y-1}~,~\,
\label{eq:sum Higgs charges}
\end{align}

\begin{align}
\mathcal{A}_{WWX}&=\mathcal{A}_{CCX}~,~\, \nonumber\\
X_{H_u} + X_{H_d} + \sum_i \left[ 3X_{q_i} + X_{l_i}\right] &= -3\left( X_{H_u}
+ X_{H_d}\right) + 18+3c~,~\, \nonumber\\
\sum_i \left[ 3X_{q_i} + X_{l_i} \right]
&=\frac{(-k_Y+\frac{11}{3})(6+c)}{k_Y-1}~,~\,
\label{eq:3Q+L}
\end{align}

\begin{align}
\mathcal{A}_{CCX}&=-3\left( X_{H_u} + X_{H_d}\right) + 18+3c\nonumber\\
&= -3 \frac{\left(k_Y-\frac{5}{3}\right)(6+c)}{k_Y-1} + 18 +3c\nonumber\\
&= 2 \frac{6+c}{k_Y-1}~.~\,
\label{eq:A_CCX}
\end{align}
Eq. (\ref{eq:A_CCX}) tells us that unless $k_Y=1$, which is inconsistent
with gauge coupling unification, we will have contributions
to the gauge anomalies from the chiral fermions. This explains why the
GS mechanism is necessary in the first place.

Assuming that all the neutrino Dirac and Majorana matrices
are holomorphic due to SUSY,
the $ia$ entry of the Dirac coupling behaves as
\begin{equation}
Y_{\mathrm{Dirac\;} ia} \sim \order{X_{H_u} + X_{l_i} + X_{\nu^c_a}}~,~\,
\end{equation}
while the $ab$ entry of the Majorana coupling goes as
\begin{align}
M_{RR\;ab} &\sim M_s \order{X_{\nu^c_a} + X_{\nu^c_b}}~.~\,
\end{align}
From the seesaw relation~\cite{Seesaw}
\begin{equation}
M_{LL} = \langle H_u \rangle^2 Y_D M_{RR}^{-1} Y_D^T~,~
\end{equation}
we find
\begin{align}
M_{LL\;ij} &\sim \frac{\langle H_u \rangle^2}{M_s^{-1}} \max_{ab} \left[
\order{X_{H_u} + X_{l_i} + X_{\nu^c_a}} \order{-X_{\nu^c_a} - X_{\nu^c_b}}
\order{X_{H_u} + X_{l_j} + X_{\nu^c_b}} \right] \nonumber\\
&\sim \frac{\langle H_u \rangle^2}{M_s^{-1}} \order{2X_{H_u} + X_{l_i} +
X_{l_j}}.
\end{align}
Combining this with the neutrino mass texture ansatz in
Eq.~(\ref{eq:neutrino}), we obtain
\begin{subequations}
\label{eq:neutrinoA}
\begin{align}
2X_{H_u} + 2X_{l_1} &= -3~,~\, \\
2X_{H_u} + 2X_{l_2} &= -5~,~\, \\
2X_{H_u} + 2X_{l_3} &= -5~.~\,
\end{align}
\end{subequations}

The only constraints on the $X$ charges of the right-handed neutrinos coming
from the previous analysis are
\begin{subequations}
\label{eq:neutrinoB}
\begin{align}
X_{H_u} + X_{l_i} + X_{\nu^c_a} &= \text{nonnegative integer}~,~\, \\
X_{\nu^c_a} + X_{\nu^c_b} &= \text{nonnegative integer}~,~\,
\end{align}
\end{subequations}
for all $i$, $a$ and $b$. This tells us that the $X$ charges of $\nu^c$ are all
odd multiples of a half and that they are all at least $5/2$. As the
right-handed neutrinos are SM singlets, they do not contribute to
$\mathcal{A}_{CCX}$, $\mathcal{A}_{WWX}$, $\mathcal{A}_{YYX}$ or
$\mathcal{A}_{YXX}$. We will only need to know their actual $X$ charges if we
wish to compute $\mathcal{A}_{XXX}$ and $\mathcal{A}_{GGX}$.

\subsection{Embedded Matter Parity}
In accordance with \cite{Dreiner:2003hw,Dreiner:2003yr,Dreiner:2006xw}, we will
assume that matter parity is embedded in the $X$ charges, although this
is not necessary in our models. The even matter
parity superfields have integral $X$ charges and the odd matter parity
superfields have $X$ charges which are odd multiples of $1/2$. The flavon and
the SUSY breaking sector all have even matter parity. This simplifies
our analysis appreciably because it prevents Higgsino-lepton and
axino-flavino-right-handed neutrino mixings. Matter parity
also protects the zero VEV of the right-handed sneutrinos
\footnote{A nonzero VEV which
is not too tiny will lead to appreciable Higgsino-lepton mixings.}. Without
it, we will have to come up with more convoluted explanations for each
of these inconvenient details. Embedding matter parity within $U(1)_X$ in this
manner circumvents problems with the nonconservation of global symmetries in
quantum gravity. By making all even matter
superfields have $X$ charges which are even, and all odd matter superfields
have $X$ charges which are odd after being multiplied by $2m$ will also give
rise
to matter parity for a positive integer $m$. This is because we can identify
matter parity with the parity of the $X$ charge multiplied by $2m$.
Eq. (\ref{eq:neutrinoA}) tells us that $m$ has to
be odd. One consequence of our previous requirement is that $X_{H_u}+X_{H_d}$
has to be an integral multiple of $\frac{1}{m}$,
and according to Eq. (\ref{eq:sum Higgs charges}), this
can only happen if $\frac{m\left(k_Y-\frac{5}{3}\right)(6+c)}{k_Y-1}$ is
integral\footnote{This constraint will definitely have to be modified if we add
additional superfields which are charged under the SM as in Section
\ref{sec:additional}.}. In addition, Eq. (\ref{eq:3Q+L}) tells us that
\begin{align}
-(X_{H_u}+ X_{H_d})&=1 \mod 3.
\end{align}
Refs.~\cite{Dreiner:2003hw,Dreiner:2003yr,Dreiner:2006xw} did not have to deal
with
this additional constraint even though they assumed that $k_Y=\frac{5}{3}$
and
$X_{H_u}+X_{H_d}=0$ because they did not always insist upon the texture given in
Eq. (\ref{eq:YE}). We present
 some possible values of $k_Y$ between $1$ and $5/3$ in Table~\ref{tab:KY-1}
where the previous two conditions are satisfied.
\begingroup
\begin{table}
\caption{Permissible values for $k_Y$ which are consistent with matter parity.}
\label{tab:KY-1}
\[
\begin{array}{|c|l|}
\hline
c&k_Y\\
\hline
0&\frac{11}{7}, \frac{7}{5}, \frac{17}{13}, \frac{5}{4}, \frac{23}{19},
\frac{13}{11}, \frac{29}{25}, \frac{8}{7}, \dots\\
1&\frac{19}{12}, \frac{47}{33}, \frac{4}{3}, \frac{65}{51}, \frac{37}{30},
\frac{83}{69}, \frac{46}{39}, \frac{101}{87}, \dots\\
2&\frac{43}{27}, \frac{13}{9}, \frac{61}{45}, \frac{35}{27}, \frac{79}{63},
\frac{11}{9}, \frac{97}{81}, \frac{53}{45}, \dots\\
3&\frac{8}{5}, \frac{19}{13}, \frac{11}{8}, \frac{25}{19}, \frac{14}{11},
\frac{31}{25}, \frac{17}{14}, \frac{37}{31}, \dots\\
\hline
\end{array}
\]
\end{table}
\endgroup

With matter parity and a huge $M_{3/2}$, the lightest SUSY particle
(LSP) is both stable and extremely heavy. This may also give us
SUSY dark matter. However, the LSP will overclose
the Universe if they were in thermal equilibrium. Thus, if the LSP
does contribute to the dark matter density, it
must be produced non-thermally.


\subsection{Solution}

We now have enough information to arrive at the  $X$ charge assignments
in Table~\ref{tab:X} with $k_Y=\frac{4}{3}$ and $c=1$.
\begingroup
\squeezetable
\begin{table}
\caption{The $X$ charge assignments for $k_Y=\frac{4}{3}$ and $c=1$ in models
without additional vector-like particles.}
\label{tab:X}
\[
\begin{array}{||c||cc|ccc|ccc|ccc|ccc|ccc|ccc||cc|c||}
\hline\hline
&H_u&H_d&Q_1&Q_2&Q_3&U^c_1&U^c_2&U^c_3&D^c_1&D^c_2&D^c_3&L_1&L_2&L_3&E^c_1&E^c_2
&E^c_3&N^c_1&N^c_2&(N^c_3)&\mathcal{A}_{CCX}&\mathcal{A}_{YXX}&k_{YX}\\
\hline\hline
A&-6&-1&\frac{11}{2}&\frac{9}{2}&\frac{5}{2}&\frac{17}{2}&\frac{11}{2}&\frac{7}{
2}&\frac{1}{2}&-\frac{1}{2}&-\frac{1}{2}&\frac{9}{2}&\frac{7}{2}&\frac{7}{2}
&\frac{3}{2}&\frac{1}{2}&-\frac{3}{2}&\geqslant\frac{5}{2}&\geqslant\frac{5}{2}
&\geqslant\frac{5}{2}&42&\frac{101}{2}&\frac{101}{84}\\
B&-5&-2&\frac{13}{2}&\frac{9}{2}&\frac{5}{2}&\frac{13}{2}&\frac{9}{2}&\frac{5}{2
}&\frac{1}{2}&\frac{1}{2}&\frac{1}{2}&\frac{7}{2}&\frac{5}{2}&\frac{5}{2}&\frac{
7}{2}&\frac{5}{2}&\frac{1}{2}&\geqslant\frac{5}{2}&\geqslant\frac{5}{2}
&\geqslant\frac{5}{2}&42&\frac{165}{2}&\frac{55}{28}\\
C&-7&0&\frac{9}{2}&\frac{9}{2}&\frac{5}{2}&\frac{21}{2}&\frac{13}{2}&\frac{9}{2}
&\frac{1}{2}&-\frac{3}{2}&-\frac{3}{2}&\frac{11}{2}&\frac{9}{2}&\frac{9}{2}
&-\frac{1}{2}&-\frac{3}{2}&-\frac{7}{2}&\geqslant\frac{5}{2}&\geqslant\frac{5}{2
}&\geqslant\frac{5}{2}&42&\frac{85}{2}&\frac{85}{84}\\
\hline\hline
\end{array}
\]
\end{table}
\endgroup
Other solutions are obtained by adding arbitrary non-negative integers
to the right-handed neutrinos' $X$ charges and adding some multiples of the
corresponding particles'
hypercharges to their $X$ charges. Only the above solutions are physical, while
the other solutions only amount to a redefinition of $X$ as hypercharge 
is itself another gauge symmetry.

Let us note that in order to have $A_{YXX}=0$, we will find irrational $U(1)_X$
charges for the SM fermions and Higgs fields since we have a quadratic equation
for $U(1)_X$ charges. $A_{YXX}=0$ can be realized for rational $U(1)_X$
charges of the SM fermions and Higgs fields if
we introduce extra vector-like particles.

\subsection{Supersymmetry Breaking}
\label{sec:SUSY breaking}
Let us assume that a chiral superfield $Z$ is responsible for the GUT-scale
SUSY breaking and that the soft SUSY breaking corrections to
the other superfields arise from their direct couplings to $Z$.
This means that the
VEV of $F$ component of $Z$ has to be non-zero,
$\langle F_Z \rangle \sim M_{\rm Pl} M_{3/2}$, where $M_{3/2}$ is comparable to
the GUT scale. While $Z$ is neutral under the SM gauge group, it may or may not
be neutral under $U(1)_X$. If $Z$ is neutral under
$U(1)_X$, it could be identified with the dilaton field $S$.

To obtain  the SM as the effective theory below the
GUT scale, one linear combination of $H_u$ and $H_d^*$
has mass around the weak scale, while the orthogonal linear
combination of $H_u$ and $H_d^*$,
 gauginos, sfermions and Higgsinos should have masses around the GUT scale.
Thus, we must make sure that these masses can be generated at
the correct scales.

If $Z$ is neutral under $U(1)_X$, we have the following K\"{a}hler potential
\begin{eqnarray}
 && \int d^4x d^2\theta d^2{\overline{\theta}}  \left(
{\frac{\overline{Z} Z}{M^2_{\rm Pl}}}  \left( H_u \overline{H_u}+
  H_d \overline{H_d} \right)
+{\frac{\overline{Z} }{M_{\rm Pl}}}
\left({\frac{\overline{A}}{M_{\rm Pl}}}\right)^{-X_{H_u}-X_{X_{H_d}}} 
H_u H_d
\right. \nonumber \\ && \left.
+{\frac{\overline{Z} Z}{M^2_{\rm Pl}}}
\left({\frac{\overline{A}}{M_{\rm Pl}}}\right)^{-X_{H_u}-X_{X_{H_d}}} H_u
H_d
 \right) + {\rm H.
C.}~,~
\end{eqnarray}
where we do not display the order one coefficients,
the third term gives us the $\mu$ term, and the rest terms
give us the scalar Higgs masses.
Then we obtain the following
``bare" scalar Higgs mass matrix:
\begin{equation}
\begin{array}{c|cc}
&H_u^\dagger & H_d\\
\hline
H_u&\mathcal{O}\left(M_{3/2}^2\right)&\mathcal{O}\left(\epsilon^{-X_{H_u}-X_{H_d
}}M_{3/2}^2\right)\\
H_d^\dagger&\mathcal{O}\left(\epsilon^{-X_{H_u}-X_{H_d}}M_{3/2}
^2\right)&\mathcal{O}\left(M_{3/2}^2\right)
\end{array}~~~.~\,
\end{equation}
Note that for $-X_{H_u}-X_{H_d}=7$, we can not have one linear combination
of $H_u$ and $H_d^*$ with mass around the weak scale.

To solve this problem, we assume that the $U(1)_X$ charge for $Z$ is equal to
$X_{H_u}+X_{H_d}$. We also assume that
$\langle F_S \rangle \sim M_{\rm Pl} M_{3/2}$. Then the K\"{a}hler potential is
given by
\begin{eqnarray}
 \int d^4x d^2\theta d^2{\overline{\theta}}
\left( \left(
{\frac{\overline{Z} Z}{M^2_{\rm Pl}}}+
 {\frac{\overline{S} S}{M^2_{\rm Pl}}}\right)
 \left( H_u \overline{H_u}+H_d \overline{H_d}\right)
+{\frac{\overline{Z}}{M_{\rm Pl}}} H_u H_d +
{\frac{\overline{Z} S}{M^2_{\rm Pl}}} H_u H_d  \right)+ {\rm H.
C.}~.~
\end{eqnarray}
And the scalar Higgs mass matrix becomes
\begin{equation}
\begin{array}{c|cc}
&H_u^\dagger & H_d\\
\hline
H_u&\mathcal{O}\left(M_{3/2}^2\right)&\mathcal{O}\left(M_{3/2}^2\right)\\
H_d^\dagger&\mathcal{O}\left(M_{3/2}
^2\right)&\mathcal{O}\left(M_{3/2}^2\right)
\end{array}~~~.~\,
\end{equation}
Thus, we can fine-tune the SUSY breaking soft masses
so that  one linear combination
of $H_u$ and $H_d^*$ has mass around the weak scale.
We can also show that the non-holomorphic contributions to the SM
Yukawa couplings can be neglected.

The gaugino masses can be generated via Eq. (\ref{SFF-gauginomass}), while
 the squark, slepton, and the sneutrino masses
can be generated via the
 K\"{a}hler potential
\begin{eqnarray}
 \int d^4x d^2\theta d^2{\overline{\theta}} \left(
{\frac{\overline{Z} Z}{M^2_{\rm Pl}}}+
 {\frac{\overline{S} S}{M^2_{\rm Pl}}}\right) \overline{\phi} \phi + {\rm
H.
C.}~,~
\end{eqnarray}
where $\phi$ denotes the SM fermion superfields.

\section{Models with Additional SM vector superfields}
\label{sec:additional}

If there is a single superfield $Z$ that breaks SUSY
and is neutral under $U(1)_X$,
to obtain one linear combination
of $H_u$ and $H_d^*$ with mass around the weak scale,
we find that
\begin{eqnarray}
X_{H_u}+X_{H_d}=2~.~
\end{eqnarray}
The soft SUSY breaking mass-squared terms for $H_u H_d$
are generated from the superpotential term $ZA^2H_u H_d/M^2_{Pl}$.

As shown above, to cancel anomalies via the GS mechanism,
we now have to introduce additional superfields
which are vector-like under the SM but not under $U(1)_X$. In this case,
 we can also have the rational solutions with $A_{YXX}=0$. However,
the price for adding these new superfields is that Eqs. (\ref{eq:anomalies})
will have to be modified and all of
the equations that follow from them, as well as the charge assignments of
Table \ref{tab:X} in general. To give a concrete example, let us introduce the
new superfields $H'_u$ and $H'_d$ with the SM charges $(1,2)_1$ and $(1,2)_{-1}$
respectively. To prevent Yukawa couplings with the SM matter superfields, we may
set their $X$ charges to some non-integral value.
The modified anomalies are now:
\begin{subequations}
\begin{align}
\mathcal{A}_{CCX}&= \sum_i \left[ 2X_{q_i} + X_{u^c_i} + X_{d^c_i} \right]~,~\,
\\
\mathcal{A}_{WWX}&= \left(X_{H_u} + X_{H_d}\right) + \left(X_{H'_u} +
X_{H'_d}\right) + \sum_i \left[ 3X_{q_i} + X_{l_i}\right]~,~\, \\
\mathcal{A}_{YYX}&= \left(X_{H_u} + X_{H_d}\right) + \left(X_{H'_u} +
X_{H'_d}\right) + \sum_i \left[ \frac{1}{3}X_{q_i} + \frac{8}{3}X_{u^c_i} +
\frac{2}{3}X_{d^c_i} + X_{l_i} +
2X_{e^c_i}\right]~,~\, \\
\mathcal{A}_{YXX}&= X_{H_u}^2 - X_{H_d}^2 + X_{H'_u}^2 - X_{H'_d}^2 +\sum_i
\left[ X_{q_i}^2 - 2X_{u^c_i}^2 + X_{d^c}^2 - X_{l_i}^2 + X_{e^c_i}^2
\right]~,~\,
\end{align}
\end{subequations}

\begin{align}
\left(k_Y-\frac{5}{3}\right)\mathcal{A}_{CCX}&=
\mathcal{A}_{WWX}+\mathcal{A}_{YYX}-\frac{8}{3}\mathcal{A}_{CCX}~,~\,
\nonumber\\
\left(k_Y-\frac{5}{3}\right) \left\{ -3\left( X_{H_u} + X_{H_d}\right) + 18+3c
\right\} &= 2\left( X_{H_u} + X_{H_d} \right) + 2\left( X_{H'_u} + X_{H'_d}
\right)~,~\, \nonumber\\
X_{H'_u} + X_{H'_d} &= -\frac{3k_Y-3}{2} \left( X_{H_u} + X_{H_d}\right) +
\frac{3k_Y-5}{2}\left(6+c\right)~.~\,
\end{align}
If $1\leqslant k_Y < 5/3$ and $X_{H_u}+X_{H_d}$ is non-negative, then
$X_{H'_u}+X_{H'_d}$ is negative. If we want the scalar $H'$ and its superpartner
to get a mass from the flavon VEV, we would also want $X_{H'_u}+X_{H'_d}$ to be
integral. Since $Z$ is neutral under $U(1)_X$,
 the H$^{\prime}$-ino mass comes from the K\"{a}hler
term $Z^\dagger {A^\dagger}^{-X_{H'_u}-X_{H'_d}} H'_u H'_d$. This is of order
$M_{3/2}\epsilon^{-X_{H'_u}-X_{H'_d}}$, which is low enough
($\sim\epsilon^4 M_{3/2}$) to
modify somewhat the gauge coupling unification. In
essence, what we have done is to shift the problem from the Higgs to
the Higgs$^{\prime}$. So, we have
\begin{align}
\mathcal{A}_{WWX}&=\mathcal{A}_{CCX}~,~\, \nonumber\\
\sum_i \left[ 3X_{q_i} + X_{l_i} \right] &= \frac{3k_Y-11}{2} \left( X_{H_u} +
X_{H_d}\right) - \frac{3k_Y-11}{2}\left(6+c\right)~.~\,
\end{align}
We present the solutions with $\mathcal{A}_{YXX}=0$
to these equations in Table~\ref{XAA}.

\begingroup
\squeezetable
\begin{table}
\caption{The $X$ charge assignments for $k_Y=4/3$, $\mathcal{A}_{YXX}=0$, and
$c=0~{\rm and}~2$ in models with $H'_u$ and $H'_d$.}
\label{XAA}
\[
\begin{array}{||c|c|c|c|c|c|c||}
\hline\hline
\text{field} & A & A & B & B & C & C \\
\hline
& c=0 & c=2 & c=0 & c=2 & c=0 & c=2\\
\hline\hline
H_u & 3a-\frac{11}{6} & 3a-{\frac {25}{6}} & 3a-\frac{5}{6} & 3a-{\frac
{19}{6}} & 3a-{\frac {17}{6}} & 3a-{\frac {31}{6}}\\
H_d & -3a+\frac{23}{6} & -3a+{\frac {37}{6}} & -3a+{\frac {17}{6}} &
-3a+{\frac {31}{6}} & -3a+{
\frac {29}{6}} & -3a+{\frac {43}{6}}\\
\hline
H'_u & \frac{169}{72} - \frac{19}{3}a  & {\frac {121}{90}}-{\frac
{107}{15}}a & {\frac {97}
{72}}-\frac{1}{3}a & -{\frac {41}{90}}-\frac{7}{3}a
 & {\frac {421}{72}}-{\frac {37}{3}}a+8a^{2} & {\frac
{463}{90}}-{\frac {179}{15}}a \\
 & + 8a^2 & +{\frac{32}{5}}a^{2} & +8a^{2} & +{\frac {32}{5}}a^{2}
& +8a^{2} & +{\frac {32}{5 }}a^{2} \\
H'_d & -\frac{457}{72} + \frac{19}{3}a  & -{\frac {571}{90}}+{\frac
{107}{15}}a & -{\frac {385}{72}}
+\frac{1}{3}a & -{\frac {409}{90}}+\frac{7}{3}a
& -{\frac {709}{72}}+{\frac {37}{3}}a & -{\frac
{913}{90}}+{\frac {179}{15}}a \\
 & -8a^2 & -{\frac {32}{5}}a^{2} & -8a^{2} & -{\frac {32}{5}}a^{2}
& -8a^{2} & -{\frac {32}{5}}a^{2} \\
\hline
q_1 & a+3 & a+3 & a+4 & a+4 & a+2 & a+2\\
q_2 & a+2 & a+2 & a+2 & a+2 & a+2 & a+2\\
q_3 & a & a & a & a & a & a\\
\hline
u^c_1 & -4a + \frac{41}{6} & -4a+\frac{55}{6} & -4a+\frac{29}{6} &
-4a+\frac{43}{6} & -4a+\frac{53}{6} & -4a+\frac{67}{6}\\
u^c_2 & -4a + \frac{23}{6} & -4a+\frac{37}{6} & -4a+\frac{17}{6} &
-4a+\frac{31}{6} & -4a+\frac{29}{6} & -4a+\frac{43}{6}\\
u^c_3 & -4a + \frac{11}{6} & -4a+{\frac {25}{6}} & -4a+\frac{5}{6} &
-4a+{\frac {19}{6}} & -4a+{\frac {17}{6}} & -4a
+{\frac {31}{6}}\\
\hline
d^c_1 & 2a-\frac{17}{6} & 2a-\frac{19}{6} & 2a-\frac{17}{6} & 2a-\frac{19}{6}
& 2a-\frac{17}{6} & 2a-\frac{19}{6}\\
d^c_2 & 2a-\frac{23}{6} & 2a-\frac{25}{6} & 2a-\frac{17}{6} & 2a-\frac{19}{6}
& 2a-\frac{29}{6} & 2a-\frac{31}{6}\\
d^c_3 & 2a-\frac{23}{6} & 2a-{\frac {25}{6}} & 2a-{\frac {17}{6}} &
2a-{\frac{19}{6}} & 2a-\frac {29}{6} & 2a-{\frac {31}{6}}\\
\hline
l_1 & -3a+\frac{1}{3} & -3a+\frac{8}{3} & -3a-\frac{2}{3} & -3a+\frac{5}{3} &
-3a+\frac{4}{3} & -3a+\frac{11}{3}\\
l_2 & -3a-\frac{2}{3} & -3a+\frac{5}{3} & -3a-\frac{5}{3} & -3a+\frac{2}{3} &
-3a+\frac{1}{3} & -3a+\frac{8}{3}\\
l_3 & -3a-\frac{2}{3} & -3a+\frac{5}{3} & -3a-\frac{5}{3} & -3a+\frac{2}{3} &
-3a+\frac{1}{3} & -3a+\frac{8}{3}\\
\hline
e^c_1 & 6a-\frac{1}{6} & 6a-\frac{17}{6} & 6a+\frac{11}{6} & 6a-\frac{5}{6} &
6a-\frac{13}{6} & 6a-\frac{29}{6}\\
e^c_2 & 6a-\frac{7}{6} & 6a-\frac{23}{6} & 6a+\frac{5}{6} & 6a-\frac{11}{6} &
6a-\frac{19}{6} & 6a-\frac{35}{6}\\
e^c_3 & 6a-\frac{19}{6} & 6a-{\frac {35}{6}}
 & 6a-\frac{7}{6} & 6a-{\frac {23}{6}} & 6a-{\frac {31}{6}} & 6a-{\frac
{47}{6}}\\
\hline
\nu^c_1 & \geqslant\frac{5}{2} & \geqslant\frac{5}{2} & \geqslant\frac{5}{2} &
\geqslant\frac{5}{2} & \geqslant\frac{5}{2} & \geqslant\frac{5}{2}\\
\nu^c_2 & \geqslant\frac{5}{2} & \geqslant\frac{5}{2} & \geqslant\frac{5}{2} &
\geqslant\frac{5}{2} & \geqslant\frac{5}{2} & \geqslant\frac{5}{2}\\
\nu^c_3 & \geqslant\frac{5}{2} & \geqslant\frac{5}{2} & \geqslant\frac{5}{2} &
\geqslant\frac{5}{2} & \geqslant\frac{5}{2} & \geqslant\frac{5}{2}\\
\hline\hline
\end{array}
\]
\end{table}
\endgroup

On the other hand, if we arrange to add full ``$SU(5)$" multiplet superfield
pairs and make their masses of order $\epsilon^4 M_{3/2}$,
the GUT scale is almost the same but the unified gauge coupling
becomes strong. Let us call the additional
superfields $XL$, $XD^c$, $\overline{XL}$ and $\overline{XD^c}$ with
the SM quantum numbers $(1,2)_{-1}$, $(\overline{3},1)_{2/3}$, $(1,2)_1$ and
$(3,1)_{-2/3}$ respectively. The anomalies are now:
\begin{subequations}
\begin{align}
\mathcal{A}_{CCX}&= \left(X_{XD^c} + X_{\overline{XD^c}}\right) + \sum_i \left[
2X_{q_i} + X_{u^c_i} + X_{d^c_i} \right]~,~\, \\
\mathcal{A}_{WWX}&= \left(X_{H_u} + X_{H_d}\right) + \left(X_{XL} +
X_{\overline{XL}}\right) + \sum_i \left[ 3X_{q_i} + X_{l_i}\right]~,~\, \\
\mathcal{A}_{YYX}&= \left(X_{H_u} + X_{H_d}\right) + \left(X_{XL} +
X_{\overline{XL}}\right) + \frac{2}{3}\left( X_{XD^c}+X_{\overline{XD^c}}\right)
+ \nonumber\\
&\quad \sum_i \left[ \frac{1}{3}X_{q_i} + \frac{8}{3}X_{u^c_i} +
\frac{2}{3}X_{d^c_i} + X_{l_i} +
2X_{e^c_i}\right]~,~\, \\
\mathcal{A}_{YXX}&= X_{H_u}^2 - X_{H_d}^2 - X_{XL}^2 + X_{\overline{XL}}^2 +
X_{XD^c}^2 - X_{\overline{XD^c}}^2 + \nonumber\\
&\quad \sum_i \left[ X_{q_i}^2 - 2X_{u^c_i}^2 + X_{d^c}^2 - X_{l_i}^2 +
X_{e^c_i}^2 \right]~,~\,
\end{align}
\end{subequations}

\begin{align}
\mathcal{A}_{CCX}&=\left(X_{XD^c} + X_{\overline{XD^c}}\right)-3\left( X_{H_u} +
X_{H_d}\right) + 18+3c ~,~\,
\end{align}

\begin{align}
\left(k_Y-\frac{5}{3}\right)\left[\left(X_{XD^c} +
X_{\overline{XD^c}}\right)-3\left( X_{H_u} + X_{H_d}\right) +
18+3c\right]\nonumber\\
= 2\left( X_{H_u} + X_{H_d} \right) + 2\left(X_{XL} + X_{\overline{XL}}\right)
-\frac{8}{3}\left(X_{XD^c} + X_{\overline{XD^c}}\right) ~,~\,
\end{align}

\begin{align}
\mathcal{A}_{WWX}&=\mathcal{A}_{CCX}~,~\, \nonumber\\
\sum_i \left[ 3X_{q_i} + X_{l_i}\right] &= -4\left( X_{H_u} + X_{H_d}\right) -
\left(X_{XL} + X_{\overline{XL}}\right) + \left(X_{XD^c} +
X_{\overline{XD^c}}\right) + 18+3c~.~\, \nonumber\\
\end{align}

To ensure that the fermionic superpartners of $XL$ and $XD^c$ have masses of the
same order of magnitude, we require that
\begin{equation}
X_{XL}+X_{\overline{XL}}=X_{XD^c}+X_{\overline{XD^c}}~,~\,
\end{equation}

\begin{equation}
X_{XL}+X_{\overline{XL}}=X_{XD^c}+X_{\overline{XD^c}}=-\frac{3k_Y-5}{k_Y-1}
\left(6+c\right)+3\left(X_{H_u}+X_{H_d}\right)~.~\,
\end{equation}
The sum is actually positive for the range of values of $k_Y$ and
$X_{H_u}+X_{H_d}$
that we are
interested in. This makes a significant difference as the superpotential
coupling $A^{X_{XL}+X_{\overline{XL}}}XL \overline{XL}$ and
$A^{X_{XL}+X_{\overline{XL}}}XD^c \overline{XD^c}$ are actually allowed provided
that the sum is an integer. Unfortunately, the suppression factor is way too
large as the sum is typically between $24$ and $33$.
We list the concrete $X$ charge assignments in Table~\ref{XII}.

\begingroup
\squeezetable
\begin{table}
\caption{The $X$ charge assignments for $k_Y=4/3$ in the models with $XL$,
$\overline{XL}$, $XD^c$, and $\overline{XD^c}$.}
\label{XII}
\[
\begin{array}{||c|ccc||}
\hline\hline
&A&B&C\\
\hline\hline
H_u&1-c&2-c&-c\\
H_d&1+c&c&2+c\\
\hline
XL&d&d&d\\
\overline{XL}&24+3c-d&24+3c-d&24+3c-d\\
\hline
XD^c&e&e&e\\
\overline{XD^c}&24+3c-e&24+3c-e&24+3c-e\\
\hline
Q_1&\frac{7}{2}&\frac{9}{2}&\frac{5}{2}\\
Q_2&\frac{5}{2}&\frac{5}{2}&\frac{5}{2}\\
Q_3&\frac{1}{2}&\frac{1}{2}&\frac{1}{2}\\
\hline
U^c_1&\frac{7}{2}+c&\frac{3}{2}+c&\frac{11}{2}+c\\
U^c_2&\frac{1}{2}+c&-\frac{1}{2}+c&\frac{3}{2}+c\\
U^c_3&-\frac{3}{2}+c&-\frac{5}{2}+c&-\frac{1}{2}+c\\
\hline
D^c_1&-\frac{1}{2}&-\frac{1}{2}&-\frac{1}{2}\\
D^c_2&-\frac{3}{2}&-\frac{1}{2}&-\frac{5}{2}\\
D^c_3&-\frac{3}{2}&-\frac{1}{2}&-\frac{5}{2}\\
\hline
L_1&-\frac{5}{2}+c&-\frac{7}{2}+c&-\frac{3}{2}+c\\
L_2&-\frac{7}{2}+c&-\frac{9}{2}+c&-\frac{5}{2}+c\\
L_3&-\frac{7}{2}+c&-\frac{9}{2}+c&-\frac{5}{2}+c\\
\hline
E^c_1&\frac{11}{2}-c&\frac{15}{2}-c&\frac{7}{2}-c\\
E^c_2&\frac{9}{2}-c&\frac{13}{2}-c&\frac{5}{2}-c\\
E^c_3&\frac{5}{2}-c&\frac{9}{2}-c&\frac{1}{2}-c\\
\hline
N^c_1&\geqslant\frac{5}{2}&\geqslant\frac{5}{2}&\geqslant\frac{5}{2}\\
N^c_2&\geqslant\frac{5}{2}&\geqslant\frac{5}{2}&\geqslant\frac{5}{2}\\
(N^c_3)&\geqslant\frac{5}{2}&\geqslant\frac{5}{2}&\geqslant\frac{5}{2}\\
\hline\hline
\mathcal{A}_{CCX}&36+6c&36+6c&36+6c\\
\hline\hline
\end{array}
\]
\end{table}
\endgroup

\section{Discussion and Conclusions}
\label{sec:conclusion}


Gauge coupling unification is a powerful tool for constructing realistic
models. We have considered the SM fermion masses and mixings in theories
obtainable from higher dimensional models, in which the SM gauge
couplings unify without invoking low energy SUSY at scales of order
$10^{16}$ GeV. An anomalous gauged flavor $U(1)_X$ symmetry plays an essential
role in our analysis. We have shown how this framework can be made
compatible with the observed fermion mass hierarchies and mixings by
employing the FN mechanism. Although low energy SUSY is absent, the
latter does play an essential role at $M_\text{GUT}$.

\begin{acknowledgments}

This work is supported in part by DOE Grant No. ~DE-FN02-91ER40626 (IG,CL,QS),
and by the
Cambridge-Mitchell Collaboration in Theoretical Cosmology (TL).

\end{acknowledgments}

\end{document}